\documentclass[aps]{revtex4}
\usepackage{graphicx}
\usepackage{eufrak}

\def\half {{1\over 2}}

\def\bea{\begin{eqnarray}}
\def\eea{\end{eqnarray}}

\def\sqr#1#2{{\vcenter{\vbox{\hrule height.#2pt
      \hbox{\vrule width.#2pt height#1pt \kern#1pt
         \vrule width.#2pt}
      \hrule height.#2pt}}}}

\def\grav {{\bf g} }

\def\figloc#1#2 {
\begin{figure}\begin{center}
    \includegraphics[width=80mm]{Fig#1.ps}
    \caption{ #2}
    \end{center}\end{figure}
}

\begin{document}
\title{ The Falling Slinky  }

\author{W. G. Unruh}
\affiliation{ CIAR Cosmology and Gravity Program\\
Dept. of Physics\\
University of B. C.\\
Vancouver, Canada V6T 1Z1\\
~
email: unruh@physics.ubc.ca}

~

~

\begin{abstract}

The slinky, released from rest hanging under its own weight, falls in a
peculiar manner. The bottom stays at rest until a wave hits it from above. Two
cases-- one unphysical one where the slinky is able to pass through itself,
and the other where the coils of the slinky collide creating a shock wave
travelling down the slinky are analysed. In the former case, the bottom begins
to move much later than in the latter. 

\end{abstract}

\maketitle
Hang a slinky up so that it is supported only on top, and let it go. The
bottom of the slinky stays at rest (does not move) for a lengthy time. Let us
analyse the behaviour in two cases, one where the slinky coils can
interpenetrate each other (a physically unrealistic situation), and one where
the coils inelastically collide. For the former case see the earlier work by Calkin
\cite{slinky} who examines the case of a spring whose equilibrium lenght is
non-zero so there is no inversion of the spring.  

Set up a labeling of the slinky with coordinate $y$ uniformly along the
slinky. The density in this coordinate is given by $\rho$ which is a constant.
Let $x$ be a vertical real space coordinate. Then the location of the point
$y$ along the slinky is given by $x(t,y)$. The stretching of the slinky will
be given by $\partial x\over \partial y$ and the force due to this stretching
is $k{\partial x\over \partial y}$. We can write the Lagrangian by 
\bea
L= \int \left[ {1\over 2}\rho \left({\partial x\over \partial t}\right)^2 -{1\over
2}k\left({\partial x\over\partial y}\right)^2 -\rho\grav x \right] dy
\eea
where $\grav$ is gravitational acceleration. This 
gives the equation of motion

\bea
\rho{\partial^2 x\over\partial t^2}-k{\partial^2x\over \partial y^2} +\rho
{\grav}
=0
\eea
Intially the slinky is supported at its top end and is stationary. The solution
is
\bea
x(y)= {1\over 2}{\rho \grav\over k} y^2.
\eea
where $y=0$ is taken to be the bottom of the slinky and $y=L$ the top. 
$k \over \rho$ is $v^2$, the wave velocity of sound (compression) waves on the
slinky.

After it is released, the boundary conditions at the two ends must be that
${\partial x\over \partial_y}=0$ at the two ends so that there are no forces
due to the stretched spring at the ends. The solution to the equations of
motion are of the form, such that  the velocity of any point on the spring is
0 at t=0 is 
\bea
x(y,t)= -{1\over 2} \grav t^2 + f(y+vt) +h(y-vt)
\eea
Ie, we have the gravitational fall of the slinky plus waves travelling to the
left and to right.
In order to have the correct boundary condition ${dx\over dt}=0$ everywhere at $t=0$ we require
\bea
{dx\over dt}\vert_{t=0} =vf'(y)-vh'(y)=0&;&0<y<L
\\
f'(y)=h'(y)&;&0<y<L
\eea
or, choosing the integration constant appropriately
\bea
f(y)=h(y); ~~~~0<y<L
\eea
In order that at all times at y=0 we have ${\partial x\over \partial y}=0$ we
require
\bea
f'(vt)+h'(-vt)=0\\
{\rm or}\\
\partial_z f(z)=\partial_z h(-z) 
\eea
Combing with $f(z)=h(z)$ for $0<z<L$ we have $f(z)=f(-z)=h(z)$ for $-L<z<L$.
Finally, demanding that ${\partial x\over \partial y}=0$ at $y=L$ gives
\bea
0=f'(L+vt)+h'(L-vt)\\
\eea
or
\bea
f(2L+z)=f(z)
\eea
Thus  $f$ and $h$ are periodic with period $2L$.

Since at t=0, we have
\bea
x(0,y)={1\over 2} \grav {y^2\over v^2}
\eea
we have
\bea
f(y)=h(y)= {\grav\over 4v^2} y^2
\eea
for $-L<y<L$ and, being periodic with period $2L$, this  determins the value at other
values of y. 

Thus, let us consider the complete solution . We can determine it
for$0< t<L/v$
by dividing the interval $0<y<L$ into two parts, $0<y<L-vt$ and $L-vt<y<L$.
For the former, we have
\bea
x(t,y)&=& -{1\over 2}gt^2 +{ \grav\over 4v^2}
\left((y-vt)^2+(y+vt)^2\right)\\
&=& -{1\over 2} \grav t^2+{ \grav\over 2v^2}(v^2 t^2 +  y^2\\
&=& {1\over 2}{ \grav\over v^2} y^2
\eea
Ie, for $y<L-vt$, the bottom of the slinky, the slinky remains static.
For the top $L>y>L-vt$, the solution is given by
\bea
x(t,y)&=& -{1\over 2}\grav t^2 + {1\over 4}{\rho \grav\over
k}\left((2L-y+vt)^2+(y+vt)^2\right)\\
&=&  { \grav\over v^2}(L^2 -Ly +{1\over 2}y^2 -Lvt)
\eea
Ie, for any point $y$ above the junction, that part of the slinky moves with
constant velocity $\grav L\over v$ after the junction moves by. The junction point, $y=L-vt$ occurs at 
\bea
X(t)={1\over 2}{\grav\over v^2}(L^2-2Lvt)
\eea

If one were to continue the solution for times longer than $L\over v$, one
would find the same behaviour, namely the slinky is divided into two, with
each section travelling at constant velocity, but with a moving boundary
(the boundary travelling at $v$ in the slinky internal coodinates). Ie, while the bottom is
stationary, the top moves with velocity ${\grav L\over v}$, then the bottom moves
with velocty $2{\grav L\over v}$,while the top continues with its former velocity, then
the top moves at $3{\grav L\over v}$, etc. iI.e., the slinky falls in steps.

Figure 1 shows the $X$ as a function $y$ and $t$ at set intervals of $t$,
showing this motion for the first complete cycle. At the end of the cycle, the
whole slinky would be falling with a uniform velocity, and the two  ends,
$y=0$ and $y=L$ have changed ends, as if the slinky were now supported at the
$y=0$ and moving with velocity $\grav L\over v$ downards.

\figloc{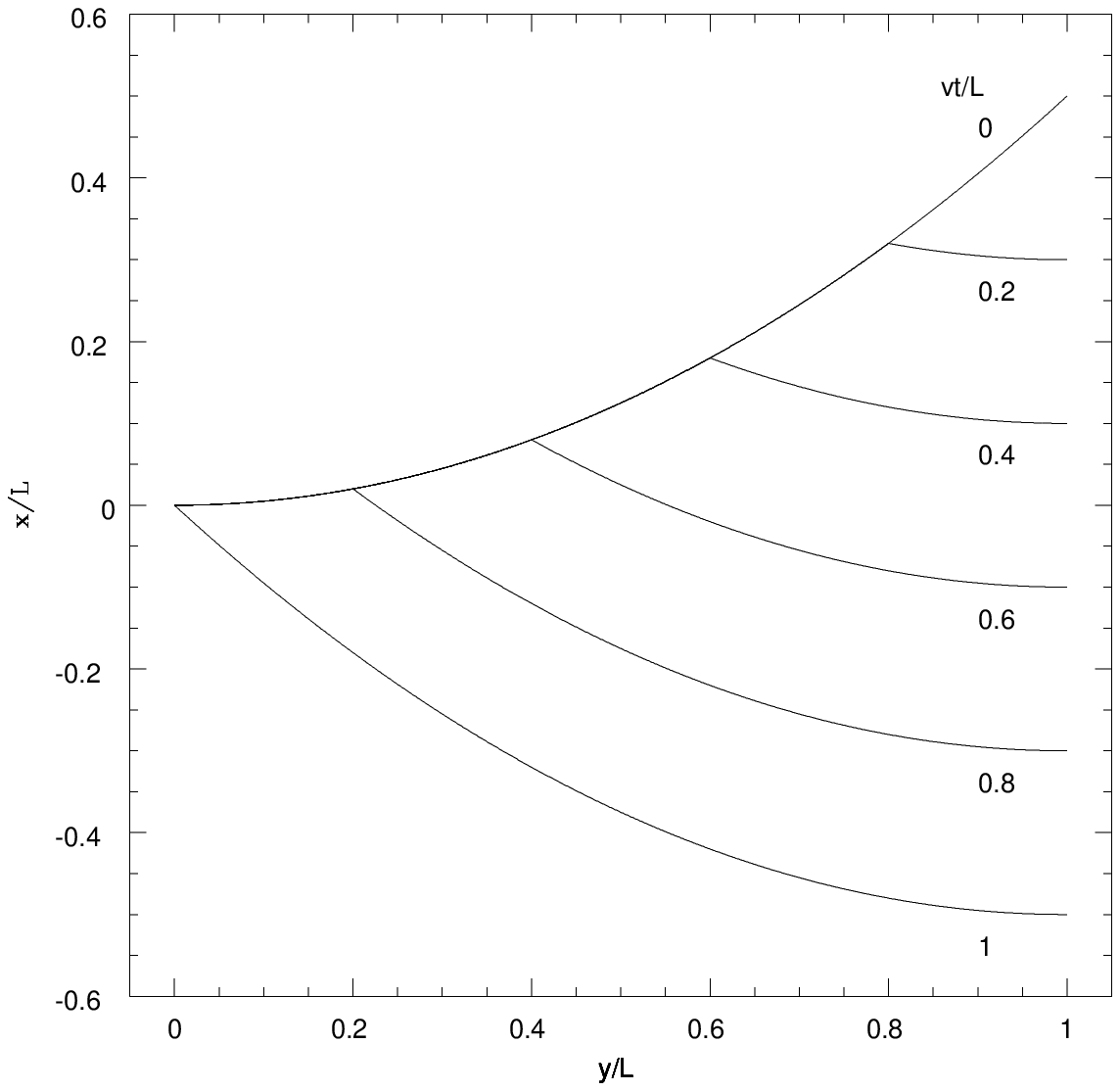}{ The location of the "yth" slinky coil at time $t$ for the
"transparent" slinky at specific times}

Of course this is all nonesense, because this solution, no matter how
interesting, assumes that one part of the slinky can interpenetrate another
part of the slinky which is not true of any slinky I know. In figure 1, we see
that after $t=0$ there are two solutions for $y$ for some values of  $x(t,y)$.
I.e., two values of "coil" coordinate, $y$ have the same location $x$. Or,
another way of phrasing it, 
at the point $y=L-vt$, the slope
$\partial x\over \partial y$ changes from positive to negative. For $y$ less
than $L-vt$ the slope is $\grav y/v^2$ while for larger values of y, it is ${\grav\over
v^2}(y-L)<0$. But a slope change from positive to negative means that the slinky has passed
through itself. Before that happens the coils will crash together, and the
above solution becomes invalid. I.e., the speed of propagation of the junction
between the parts of the slinky is NOT equal to speed of sound along the
slinky.

Let us assume that, when the coils come together the collision is a perfectly
inelastic collision. Let us assume that above some point $Y(t)$ the coils are
all together, while below that point, the slinky, as above, remains
motionless. Then we have a mass $M=\rho (L-Y)$ above that transition point,
while below it, we assume as above that the slinky remains motionless. Note
that if ${dY\over dt}<v$ this will clearly not be a valid solution. 

The velocity in $x$-space of that point $Y(t)$ is 
\bea
U= {dY\over dt}{\partial x\over\partial y}\vert_{y=Y}
\eea
Thus the Momentum equation for that mass above the transition point is 
\bea
{dMU\over dt}= -M\grav -k{\partial x\over \partial y}\vert_{y=Y}
\eea
or
\bea
{d\left(\rho (L-Y) {dY\over dt} {\rho \grav \over k}Y\right) \over dt}= -\rho \grav (L-Y)-k({\rho
\grav\over k} Y) = -\rho \grav L
\eea
The solution is
\bea
{1\over v^2} (L-Y)^2({Y\over 3L}+{1\over 6})= {1\over 2} t^2+ct
\eea
If $c\neq 0$, then for small $t$ we have $y=L-v\sqrt{ct}$, and the velocity
$dY\over dt$ goes to infinity as $t\rightarrow 0$. Thus we have
\bea
(L-Y)^2({2Y\over 3L} +{1\over 3})= v^2 t^2
 \eea

\figloc{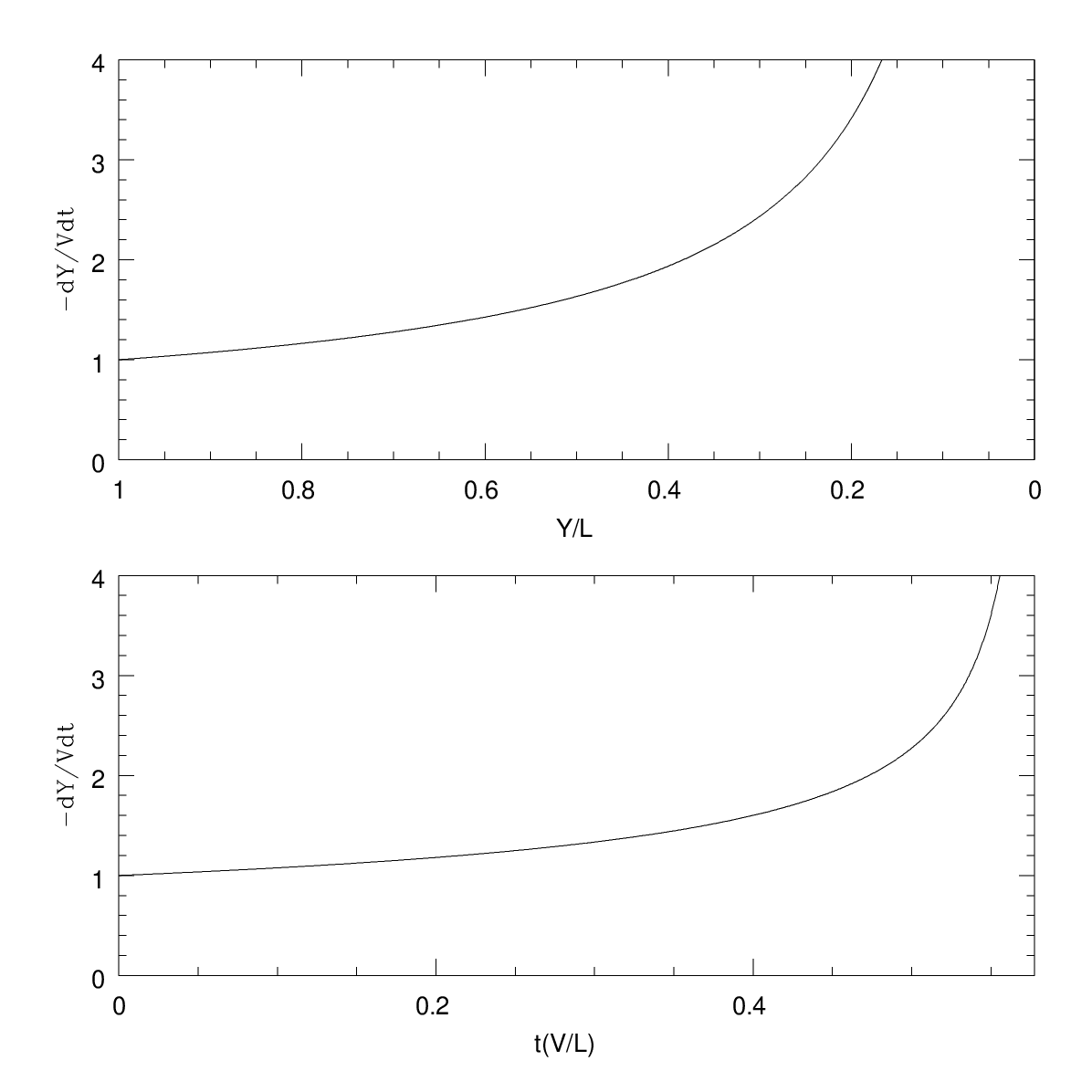}{-$dY\over dt$ as a function of $Y$ and $t$ for the boundary of the
shock wave in the falling slinky.}

As we can see from figure 2, $-{dY\over dt}$ is always greater than $v$. and
approaches infinity as $Y\rightarrow 0$. This falling slinky has a shock wave
where the top collapsed part of the slinky crashes into the lower stationary
part, and as is usual
for shock waves, they travel faster than the velocity of sound (in this case
$v$) in the medium. In physical space, 
\bea
{dX\over dt} = {dY\over dt}{\partial y\over \partial x}\vert_{y=Y} = {v^2L t\over
L-Y}{\grav\over v^2} = {\grav t}{L\over L-Y}
\eea
which goes to the finite velocity $\grav t=\grav({L\over \sqrt{3}v})$ as $Y\rightarrow 0$.

\figloc{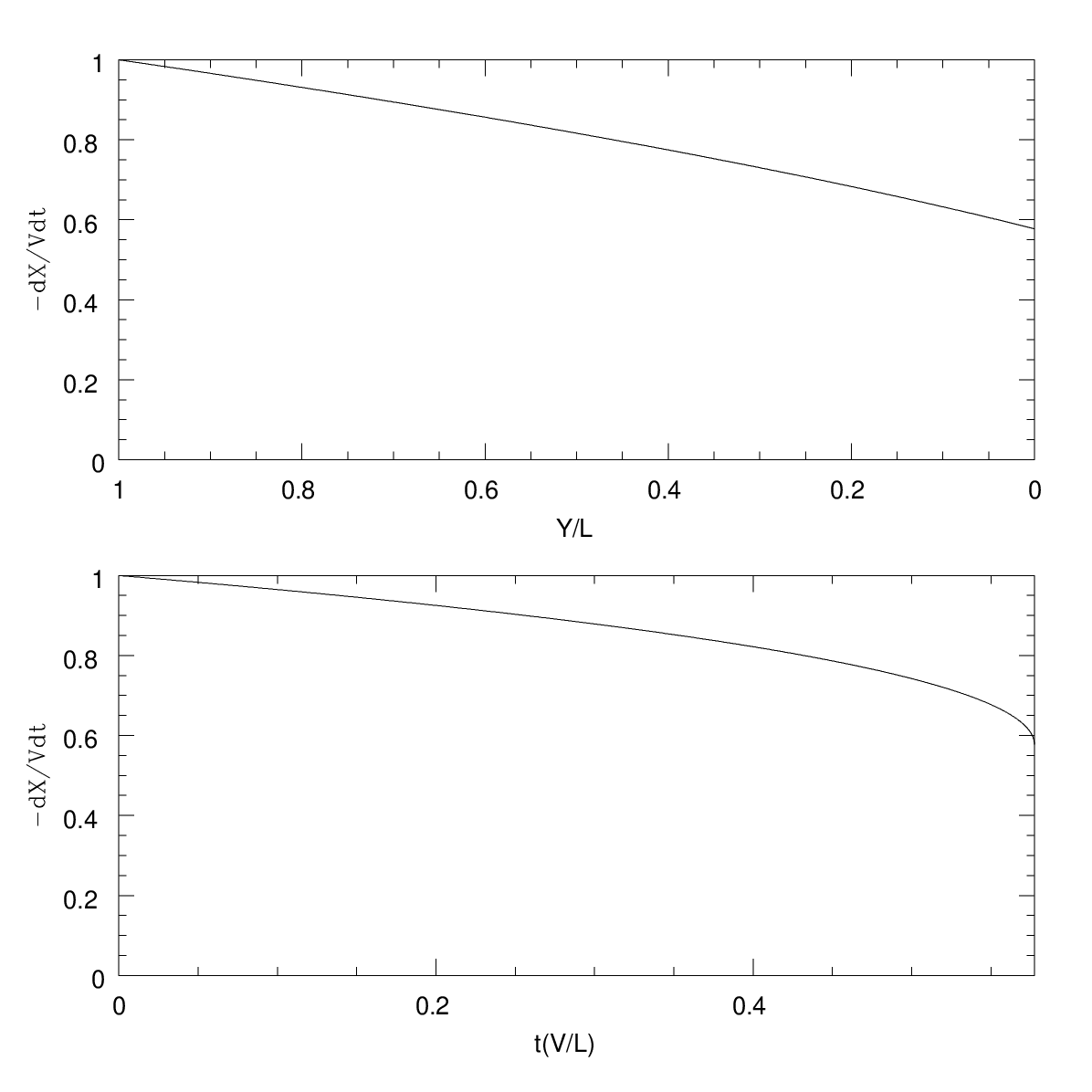}{The physical velocity, -$dX\over dt$, as a function of Y and t for the boundary of the
shock wave in the falling slinky.}
In figure 3 I have plotted $-{v^2\over \grav L}{dX\over dt}$ as a function of
${Y\over L}$ and $vt/L$
( which is $1/\sqrt{3}$ when $Y(t)=0$.)

Note that initially the physical velocity of the shock front is  finite
($gL\over v$) and
is not just proportional to $\grav t$, the free fall velocity. It then falls to
$\grav t= {\grav\over\sqrt{3}}{L\over v}$ as the shock approaches the bottom end of
the slinky.

There is a simpler way of deriving the equation for the shock wave location.
The center of mass of the coil falls with a position of $CM=CM_0 -{\half}\grav
t^2)$ where $CM$  is the center of mass of the coil
Assuming that the bottom of the coil remains stationary
\bea
CM= {1\over L}((L-Y(t)) X(t) + \int_0^Y(t) x dy=(L-Y(t))({g\over 2v^2}Y(t)^2
+{g\over 3v^2} Y(t)^3)= CM(0) -{g\over 2}t^2
\eea
or
\bea
-(L-Y(t))^2 {1\over 3}(2Y(t)+L)= v^2 t^2 L
\eea
or
\bea
{dY(t)\over dt}= -{v\sqrt{(2Y(t)+L)L/3}\over Y(t)}>v
\eea
for $0<Y<L$ which is indentical to the above conservation of momentum
argument, and shows that the shock front $Y(t)$ always travels at a velocity
greater than the velocity of sound in $y$ coordinates.

The total energy of the spring, potential plus kinetic is
\bea 
E&=&M(t){dX(t)\over dt}^2 +M(t)\grav X(t)+\int_0^{Y(t)} \rho \grav x(y)+{1\over 2} k\left({\partial x\over
\partial y}\right)^2 dy 
\eea
The first term is the kinetic energy of the collapsed part of the slinky at
the top. The second is the gravitational potential energy of that collapsed
part. The third the gravitational potential energy of the uncollapsed spring,
and the fourth is the potential energy of the expanded spring. 

Using the expressions for the velocity, $U(t)= {dX\over dt}={dX\over dY}
{dY\over dt}$, the velocity of the shock in Lagrangian coordinates ${dY\over
dt}= -v {\sqrt{(2Y+L\over 3L}\over L}$ the
collapsed mass, $M=\rho (L-Y)$, and the potential energy of the expanded
spring I get
\bea
E&=& \left[(L-Y(t))(U(t)^2+\grav ({\grav\over 2v^2}Y(t))+{\grav^2\over
6v^2}Y(t)^3+{1\over 2}v^2 ({\grav \over v^2})^2 {Y(t)^3\over 3}\right]
\\
&=& \rho {\grav^2 \over 6 v^2 L^3} (L^3+L^2Y+LY^2-Y^3)
  \eea
The total energy drops by a factor of 2 as the shock wave travels from the
top to the bottom of the spring, but  this drop is not linear in either $Y$ or $t$. The
energy loss of course occurs in the collision between the coils of the slinky,
converting from mechanical energy to heat energy in the coils.

Figure 4 are plots of the total energy as a function of $Y$ and of $t$.
The dotted curve is the  potential energy in the stretched spring
remaining as the shock reaches the point $Y$ plus the gravitational potential
energy at time $t=0$ plus the initial gravitational potential energy of the
center of mass of the spring. The center of mass energy we expect to be conserved.
That it is the internal potential energy of the
stretched spring that is converted to heat is clear if one goes into the center of
mass frame of the spring which falls with acceleration $\grav$.
 There the only potential energy is the stretched
spring, and at the end, the stretched spring is at rest and unstretched, with no internal
motion. 

\figloc{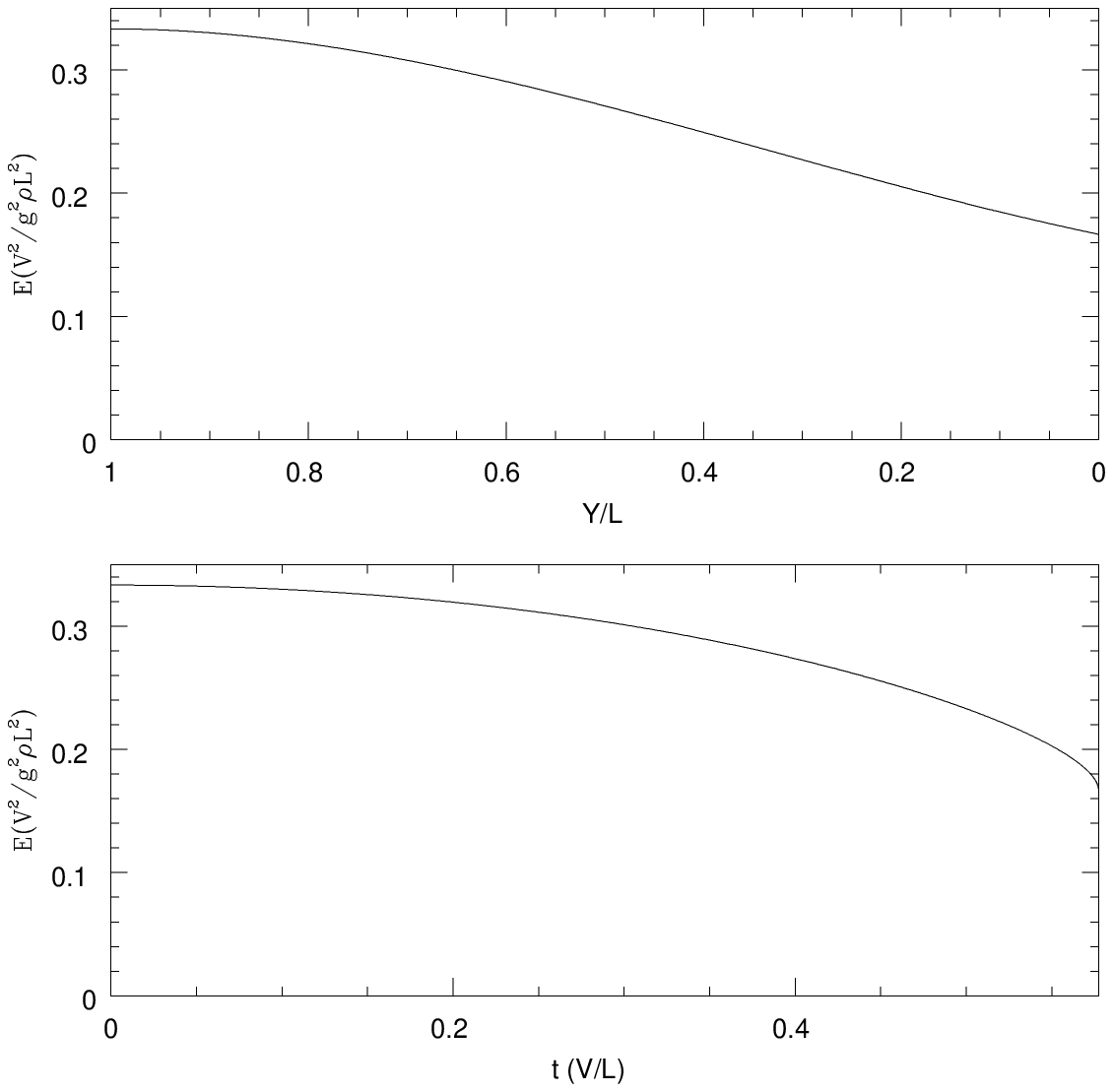}{The total energy (Kinetic plus potential energy--both gravitational
with the bottom of the slinky as potential energy reference point and
potential energy in the spring) as a function of $Y$ and of $t$. Note that in
neither case is the loss of energy linear. In fact, it is at the end of the
trajectory of the shock that the largest rate of loss of energy occurs, where
both the graviational and  spring potential energies are the least.}

This problem illustrates a number of features of shock fronts\cite{shock}. They do not
conserve energy in their gross motion (of course the lost energy goes into
heating the system behind the shock). The motion of the shock front is also
larger than the velocity of sound in the medium (in this case it goes to
infinity in the $y$ coordinates, while the velocity of `sound', $v$, is a constant).
This problem does differ from many other shocks in that the heating due to the
energy dissipation in the shock does not affect the equations of motion of the
shock itself.

If the equilibrium of the spring is not at ${dx\over dy}=0$, but has some
equilibrium value $a$, then the static solution for the hanging spring is
\bea
x={\grav \rho\over 2 k}(y+a)^2
\eea
Time dependent solution is again
\bea
x(t,y)= -{1\over 2} \grav t^2 +f(y-vt)+g(y+vt)
\eea
with initial condition again 
\bea
x(0,y)= {\grav\rho\over 2 k}(y+a)^2
\\
\partial_t x(0,y)=0
\eea
giving $f(y)=h(y)$
The boundary conditions at $y=0$ and $y=L$ are 
\bea
\partial_y x(t,0)=a \\
\partial_y x(t,L)=a
\eea
The solution is 
\bea
f(y)= \left\{ \begin{array}{ll}{\grav\over v^2}(y+a)^2 & -L<y<L\\
                          {\grav\over v^2}((y-2L)+a)^2 +4aL& L<y<3L
		\end{array}
			\right.
\eea
Thus the solution for $0<t<{L\over v}$ is 
\bea
x(t,y) = \left\{\begin{array}{ll}{ \grav\over 2v^2}(y+a)^2 & 0<y<vt\\
                          { \grav\over 2v^2}(y-L+a)^2+L^2-2Lvt & vt<y<L 
                          
		\end{array}\right.
\eea
The condition that the spring elements not collide is that ${dx\over dy}$ not
be zero anywhere for $0<y<L$ which is that $y-L+a$ not be zero anywhere. This
implies that $a>L$ is the condition that a freely suspended spring not have a
zero slope anywhere. The stretch at the top of the freely suspended spring is
${\grav\over v^2}(L+a)$ while at the bottom it is ${\grav\over v^2}a$ which has
a ratio of $ {L\over a}+1$. Thus the condition that the coils of the spring
not collide (as in ref \cite{slinky} when the freely suspended spring is dropped is that the ratio of
the stretch at the top and the bottom must be less than a factor of 2 ($a>L$)
when the spring is suspended by its end.

\end{document}